\documentclass[aps,prd,preprint,nofootinbib]{revtex4}
\usepackage{amsfonts}
\usepackage{mathrsfs}
\usepackage{graphicx}
\usepackage{amsmath}
\usepackage{amssymb}
\usepackage{multirow}
\usepackage{subfigure}
\usepackage{epsfig}
\usepackage{graphicx}
\usepackage{booktabs}
\usepackage{array}
\usepackage{tabularx}
\allowdisplaybreaks[1]
\graphicspath{{fig/}{dia/}} \DeclareGraphicsExtensions{.eps}
\begin{document}
\baselineskip 20pt
\title{NLO QCD Corrections to Inclusive Charmonium and $B_c$ Meson Production in $W^+$ Decays\\ [0.7cm]}

\author{Zi-Qiang Chen$^1$\footnote[2]{chenziqiang13@mails.ucas.ac.cn}, Hao Yang$^1$\footnote[3]{yanghao174@mails.ucas.ac.cn} and Cong-Feng Qiao$^{1,2}$\footnote[1]{qiaocf@ucas.ac.cn, corresponding author}}
\affiliation{$^1$ School of Physics, University of Chinese Academy of Science, Yuquan Road 19A, Beijing 10049 \\
$^2$ CAS Key Laboratory of Vacuum Physics, Beijing 100049, China}
\author{~\\}

\begin{abstract}
~\\ [-0.3cm]
We calculate the next-to-leading order (NLO) quantum chromodynamics (QCD) corrections to inclusive processes $W^+\to J/\psi(\eta_c)+c+\bar{s}+X$ and $W^+\to B_c(B_c^{*})+b+\bar{s}+X$ in the framework of nonrelativistic QCD (NRQCD) factorization formalism. Result indicates that the NLO corrections are significant, and the uncertainties in theoretical predictions with NLO corrections are greatly reduced. The charmonium and $B_c$ meson yielding rates at the Large Hadron Collider (LHC) are given.

\vspace {7mm} \noindent {PACS numbers: 12.38.Bx, 12.39.Jh, 13.38.Be, 14.40.Pq}

\end{abstract}
\maketitle

\section{Introduction}

In the standard model (SM), the $W$ boson mass is generated through the electroweak spontaneous breaking mechanism.
Precise measurement of $W$ boson mass and its decay width turns out to be a unique test of the SM and hence a probe for new physics.
At the Large Hadron Collider (LHC), a huge number of $W$ bosons are produced and recorded, which enables the research on $W$ physics feasible and meaningful.

Heavy quarkonium and as well $B_c$ meson production keeps on being an interesting and hot topic to study in high energy physics for decades, which may enrich our knowledge on the properties of quarkonium and the nature of perturbative QCD. Note, hereafter for simplicity the $B_c$ respresents for both scalar $B_c$ and vector $B_c^{*}$ unless specifically mentioned.
Based on the nonrelativistic QCD (NRQCD) factorization formalism \cite{nrqcd}, direct hadroproduction of quarkonium and $B_c$ meson was studied extensively \cite{hdpro1,hdpro2,hdpro3,hdpro4,hdpro5,hdpro6,hdpro7,hdpro8,hdpro9,hdpro10,hdpro11}.
In addition to the direct production, indirect production also stands as an independent and important source for those double-heavy measons.
The quarkonium and $B_c$ meson production through top quark \cite{topdecay} and $Z_0$ decays \cite{Zdecay1,Zdecay2,Zdecay3} had been investigated at up to the next-to-leading order (NLO) accuracy.
For indirect quarkonium and $B_c$ in $W$ decays, the leading order (LO) analyses were performed in Refs.\cite{WdecayLO1,WdecayLO2}.
It turned out that the theoretical uncertainties at LO are very large, which suggests, and was partly confirmed, 
that the higher order QCD corrections in charmonium and $B_c$ productions are usually very important, even crucial sometimes, for the sake of phenomenological use. To this end, we calculate in this work the NLO QCD corrections to the inclusive charmonium and $B_c$ production in $W^+$ decays.

The rest of the paper is organized as follows. In section \uppercase\expandafter{\romannumeral2} we present the LO  calculation of $W^+$ decay to charmonium and $B_c$ mesons.
In section \uppercase\expandafter{\romannumeral3}, some technical details in the calculation of NLO corrections are given.
In section \uppercase\expandafter{\romannumeral4}, the numerical evaluation for concerned processes is performed at NLO QCD accuracy.
The last section is remained for summary and conclusions.

\section{The LO decay width}

\begin{figure}
	\centering
	\includegraphics[scale=0.42]{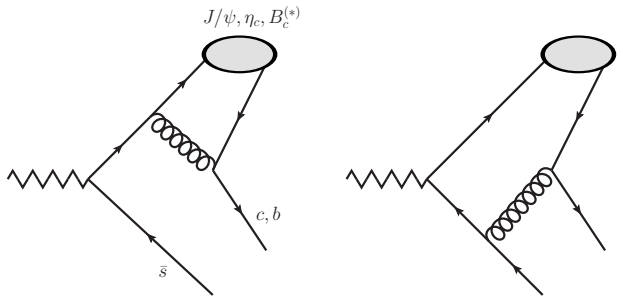}
		\caption{The LO Feynman diagrams for charmonium and $B_c$ meson production in $W^+$ decays.}
	\label{fig_tree}
\end{figure}

At the LO in $\alpha_s$, inclusive charmonium and $B_c$ meson production through $W^+$ decays are described by the processes
\begin{align}
&W^+(p_W)\to J/\psi(\eta_c)(p_H)+c(p_Q)+\bar{s}(p_s), \nonumber \\
&W^+(p_W)\to B_c(B_c^{*})(p_H)+b(p_Q)+\bar{s}(p_s)
\end{align}
as shown in Fig.\ref{fig_tree}.
The initial and final state particles are on the mass shell: $p_W^2=m_W^2$, $p_H^2=m_H^2$, $p_Q^2=m_Q^2$ and $p_s^2=0$.
We also introduce the Mandelstam variables: $s_1=(p_H+p_Q)^2$, $s_2=(p_H+p_s)^2$.
The CKM-suppressed processes, such as $W^+\to B_c(B_c^{*})+c+\bar{c}$, are not included in our calculation.
The amplitudes of these processes are suppressed at least by a Wolfenstein parameter $\lambda$.
Taking $\lambda \sim \alpha_s(2m_c)$, the suppress factor for decay width can be estimated as $\mathcal{O}(\alpha_s^2)$, which means that the contribution from these processes are less significant than the NLO corrections.

At the leading order of the relative velocity expansion, it is legitimate to take $p_c=p_{\bar{c}}$, $m_H=2m_c$ for charmonium production and $p_c=\frac{m_c}{m_b}p_{\bar{b}}$, $m_H=m_b+m_c$ for $B_c$ production. The spin projection operator has the form
\begin{equation}
\Pi(n)=\frac{1}{2\sqrt{m_H}}\epsilon(n)(p_H\!\!\!\!\!\!\!/\ \ +m_H)\otimes \left(\frac{1_c}{\sqrt{N_c}}\right),
\end{equation}
where $\epsilon(^1S_0)=\gamma_5$ and $\epsilon(^3S_1)=\epsilon \!\!\!/$.
The decay width at LO reads:
\begin{equation}
\Gamma_{\rm born}=\frac{|\Psi_H(0)|^2}{2m_W}\frac{1}{3}\int\sum |\mathcal{M}_{\rm born}|^2d{\rm PS}_3.
\end{equation}
Here, $\sum$ sums over the polarizations and colors of the initial and final particles, $\frac{1}{3}$ comes from the spin average of the initial $W^+$ boson, $d{\rm PS}_3$ stands for the three-body phase space, which can be expressed as
\begin{equation}
\int d{\rm PS}_3=\frac{1}{32\pi^3}\int_{m_H}^{E_H^+}dE_H\int_{E_s^{\rm -}}^{E_s^{\rm +}}dE_s,
\end{equation}
in the rest frame of $W^+$. Here, $E_H$ and $E_s$ represent the energy of final state hadron and $s$ quark respectively.
The upper and lower bounds of above integration are
\begin{align}
&E_H^+=\frac{m_W^2+m_H^2-m_Q^2}{2m_W}, \nonumber \\
&E_s^{\rm \pm}=\frac{1}{2}\left(1-\frac{m_Q^2}{M_{Qs}^2}\right)\left(m_W-E_H\pm\sqrt{E_H^2-m_H^2}\right).
\end{align}
with
\begin{equation}
M_{Qs}^2=m_W^2+m_H^2-2m_WE_H.
\end{equation}

\section{The NLO corrections}

At the NLO, the $W^+$ boson decay to charmonium and $B_c(B_c^{*})$ meson include the virtual and real corrections of $W^+\to J/\psi(\eta_c)+c+\bar{s}$ and $W^+\to B_c(B_c^{*})+b+\bar{s}$ processes.
For $\eta_c$ production, new subprocess $W^+\to \eta_c+u+\bar{d}+g$ should also be included.
In the computation of NLO corrections, the conventional dimensional regularization with $D=4-2\epsilon$ is adopted to regularize the ultraviolet (UV) and infrared (IR) divergences.
The method proposed in \cite{GA5a,GA5b} is used to deal with the $D$ dimensions $\gamma_5$ trace.

In the calculation, the package FeynArts \cite{feynarts} is used to generate Feynman diagrams; FeynCalc \cite{feyncalc1,feyncalc2} and FORM \cite{form1,form2} are used to perform algebraic calculation; FIRE \cite{fire1,fire2} is employed to reduce the Feynman integrals into the master integrals ($A_0,B_0,C_0,D_0$); With the help of Ref.\cite{scalarint} and Package-X \cite{packagex}, the master integrals are calculated analytically, and the results are checked by LoopTools \cite{looptools};
The numerical phase space integration is performed by CUBA\cite{cuba}.

\subsection{Virtual corrections}

\begin{figure}
	\centering
	\includegraphics[scale=0.42]{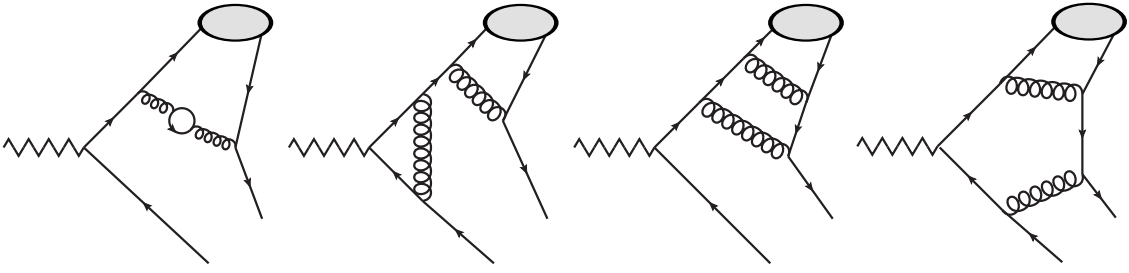}
		\caption{Typical Feynman diagrams in virtual corrections.}
	\label{fig_vdia}
\end{figure}

Typical Feynman diagrams in virtual corrections are shown in Fig.\ref{fig_vdia}.
The contribution from virtual corrections can be formulated as
\begin{equation}
\Gamma_{\rm virtual}=\frac{|\Psi_H(0)|^2}{2m_W}\frac{1}{3}\int\sum 2{\rm Re}(\mathcal{M}_{\rm virtual}\mathcal{M}_{\rm born}^*)d{\rm PS}_3.
\end{equation}
Here, ${\rm Re}(\mathcal{M}_{\rm virtual}\mathcal{M}_{\rm born}^*)$ contains both UV and IR singularities.
Since we set $p_c=p_{\bar c}$ and $p_c=\frac{m_c}{m_b}p_{\bar{b}}$ before the calculation of Feynman integrals, the Coulomb singularity are not expected to appear in our calculation \cite{thresholdexpansion}.

The UV singularities are removed by renormalization.
For the renormalization of heavy quark field ($Z_2$), heavy quark mass ($Z_m$) and light quark field ($Z_l$), we take the on-shell (OS) scheme; for the renormalization of gluon filed ($Z_3$) and strong coupling constant ($Z_g$), the modified minimal-subtraction ($\overline{\rm MS}$) schemes are used:
\begin{align}
\delta Z_2^{\rm OS}=&-C_F\frac{\alpha_s}{4\pi}
\left[\frac{1}{\epsilon_{\rm UV}}+\frac{2}{\epsilon_{\rm IR}}
-3\gamma_E+3\ln\frac{4\pi\mu^2}{m_Q^2}+4\right],
\nonumber\\
 \delta Z_m^{\rm OS}=&-3C_F\frac{\alpha_s}{4\pi}
\left[\frac{1}{\epsilon_{\rm
UV}}-\gamma_E+\ln\frac{4\pi\mu^2}{m_Q^2} +\frac{4}{3}\right], \nonumber\\
 \delta Z_l^{\rm OS}=&-C_F\frac{\alpha_s}{4\pi}
\left[\frac{1}{\epsilon_{\rm UV}}-\frac{1}{\epsilon_{\rm IR}}\right], \nonumber\\
\delta Z_3^{\overline{\rm MS}}=&\dfrac{\alpha_s}{4\pi}(\beta_0-2C_A)\left[\dfrac{1}{\epsilon_{\rm UV}} -\gamma_E +\ln(4\pi) \right],
 \nonumber\\
  \delta Z_g^{\overline{\rm MS}}=&-\frac{\beta_0}{2}\,
  \frac{\alpha_s}{4\pi}
  \left[\frac{1}{\epsilon_{\rm UV}} -\gamma_E + \ln(4\pi)
  \right].
\end{align}
Here, $\mu$ is the renormalization scale, $\gamma_E$ is the Euler's constant; $\beta_0=(11/3)C_A-(4/3)T_fn_f$ is the one-loop coefficient of QCD beta function, $n_f$ is the number of active quarks; $C_A=3$, $C_F=4/3$ and $T_F=1/2$ are color factors.
Note, final result is independent of $\delta Z_3$, because terms proportional to $\delta Z_3$ from vertex correction cancel with that from selfenergy correction.

In virtual corrections, the IR singularities arise when the gluon connecting two on shell partons is soft or collinear to final $\bar{s}$ quark. Due to $p_c=p_{\bar c}$ or $p_c=\frac{m_c}{m_b}p_{\bar{b}}$, parts of IR singularities cancel each other \cite{nrqcd}.
The remaining are canceled by the real corrections according to the Kinoshita-Lee-Nauenberg theorem \cite{KLN1,KLN2}.

\subsection{Real corrections}

\begin{figure}
	\centering
	\includegraphics[scale=0.42]{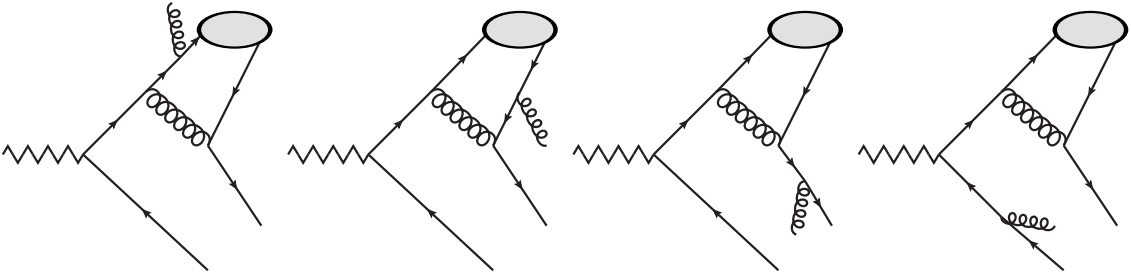}
		\caption{Typical Feynman diagrams in real corrections.}
	\label{fig_rdia}
\end{figure}

Typical Feynman diagrams in real corrections are shown in Fig.\ref{fig_rdia}.
In the calculation of the real corrections, the phase space slicing method \cite{twocutoff} is adopted to separate the IR singularities.
By introducing soft cut $\delta_s$ and collinear cut $\delta_c$, the phase space can be separated into three regions:
\begin{itemize}
\item Soft: $p_g^0<\frac{m_W}{2}\delta_s$;
\item Hard collinear: $p_g^0>\frac{m_W}{2}\delta_s$, $M_{sg}^2<m_W^2\delta_c$;
\item Hard non-collinear: $p_g^0>\frac{m_W}{2}\delta_s$, $M_{sg}^2>m_W^2\delta_c$.
\end{itemize}
Here, $M_{sg}=(p_s+p_g)^2$ is the invariant mass of $\bar{s}$ and $g$ system.
Then the real corrections can be written as
\begin{equation}
\Gamma_{\rm real}=\Gamma_{\rm real}^{\rm S}+\Gamma_{\rm real}^{\rm HC}+\Gamma_{\rm real}^{\rm HNC},
\end{equation}
where the superscripts ``S'', ``HC'', ``HNC'' represent the ``soft'', ``hard collinear'', ``hard non-collinear'' region respectively.

According to Ref.\cite{twocutoff}, the contributions from soft part and hard collinear part reads
\begin{align}
\Gamma_{\rm real}^{\rm S}=&\Gamma_{\rm born}C_F\frac{\alpha_s}{2\pi}\frac{\Gamma(1-\epsilon)}{\Gamma(1-2\epsilon)}\left(\frac{4\pi\mu^2}{m_W^2}\right)^\epsilon\frac{1}{\delta_s^{2\epsilon}}\nonumber \\
\times&\bigg[\frac{1}{\epsilon^2}+\frac{1}{\epsilon}\left(1-\ln\frac{m_W^2}{m_Q^2}-2\ln \frac{m_H^2+m_W^2-s_1-s_2}{m_W^2-s_1}\right)+{\rm finite}\bigg], \nonumber \\
\Gamma_{\rm real}^{\rm HC}=&\Gamma_{\rm born}C_F\frac{\alpha_s}{2\pi}\frac{\Gamma(1-\epsilon)}{\Gamma(1-2\epsilon)}\left(\frac{4\pi\mu^2}{m_W^2}\right)\nonumber \\
\times&\bigg[\frac{1}{\epsilon}\left(2\ln\frac{m_W^2\delta_s}{m_W^2-s_1}+\frac{3}{2}\right)-\frac{3}{2}\ln\delta_c-2\ln\delta_c\ln\frac{m_W^2\delta_s}{m_W^2-s_1}\nonumber \\
-&\left(\ln\frac{m_W^2\delta_s}{m_W^2-s_1}\right)^2-\frac{2\pi^2-21}{6}\bigg].
\end{align}
In the case of hard non-collinear part, the decay width reads
\begin{equation}
\Gamma_{\rm real}^{\rm HNC}=\frac{|\Psi_H(0)|^2}{2m_W}\frac{1}{3}\int^{\rm HNC}\sum |\mathcal{M}_{\rm real}|^2d{\rm PS}_4,
\end{equation}
where the four-body phase space $d{\rm PS}_4$ with cut can be written as
\begin{align}
\int^{\rm HNC}d{\rm PS}_4=\frac{1}{512\pi^6}&\int_{m_H}^{E_H^+}dE_H\int_{\sqrt{\delta_c}m_W}^{M_{sg}^+}dM_{sg}\int_{E_{sg}^-}^{E_{sg}^+}dE_{sg}\frac{M_{sg}}{\sqrt{E_{sg}^2-M_{sg}^2}}\nonumber \\
&\int_{{\rm max}(E_g^-,\delta_sm_W/2)}^{E_g^+}dE_g\Theta(E_g^+-\delta_sm_W/2)\int_0^{2\pi}d\eta_{sg},
\end{align}
with
\begin{align}
&E_H^+=\frac{m_W^2+m_H^2-(m_Q+\sqrt{\delta_c}m_W)^2}{2m_W},\nonumber \\
&M_{csg}=\sqrt{m_W^2+m_H^2-2m_WE_\psi},\nonumber \\
&M_{sg}^+=M_{csg}-m_Q,\nonumber \\
&E_{sg}^\pm = \frac{1}{4m_WM_{csg}^2}\bigg[(M_{csg}^2-m_Q^2+M_{sg}^2)(m_W^2-m_H^2+M_{csg}^2)\nonumber \\
&\quad\quad\quad \pm \sqrt{\lambda(M_{csg}^2,m_Q^2,M_{sg}^2)\lambda(m_W^2,m_H^2,M_{csg}^2)}\bigg],\nonumber \\
&E_g^\pm=\frac{1}{2}\left(E_{sg}\pm\sqrt{E_{sg}^2-M_{sg}^2}\right),
\end{align}
where $\lambda(s,m_a^2,m_b^2)=[s-(m_a+m_b)^2][s-(m_a-m_b)^2]$, and $\Theta(x)$ is the unit step function which return 1 when $x>0$ and 0 for other case.
After summing up these three parts, their dependence on technical cut are eliminated as expected.

\begin{figure}
	\centering
	\includegraphics[scale=0.42]{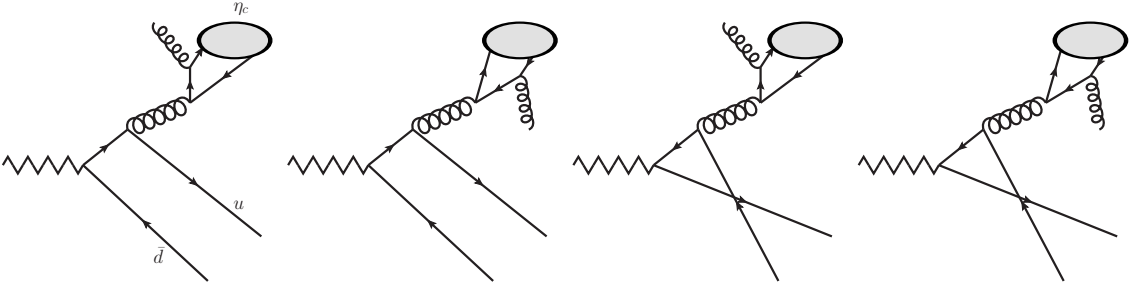}
		\caption{Feynman diagrams of $W^+\to \eta_c+u+\bar{d}+g$ process.}
	\label{fig_newpro}
\end{figure}

For the subprocess $W^+\to \eta_c+u+\bar{d}+g$, there are 4 diagrams, as shown in Fig.\ref{fig_newpro}.
The IR singularities are eliminated after summing all the amplitude square parts. The decay width can be calculated directly in 4 dimensions as $\Gamma_{\rm real}^{\rm HNC}$.

\section{Numerical results}

For the numerical calculation, following input parameters are used
\begin{align}
&m_W=80.399 {\rm GeV},\quad m_c=1.5\pm 0.1 {\rm GeV},\quad m_b=4.9\pm 0.2 {\rm GeV},\quad \alpha=1/137.065,\nonumber \\
&\quad\quad {\rm sin}^2\theta_W=0.2312,\quad|\Psi_{J/\psi}^{\rm LO}(0)|^2=\frac{0.528}{4\pi} {\rm GeV}^3,\quad |\Psi_{J/\psi}^{\rm NLO}(0)|^2=\frac{0.903}{4\pi} {\rm GeV}^3,\nonumber \\
&\quad\quad\quad\quad \Psi_{\eta_c}(0)=\Psi_{J/\psi}(0),\quad |\Psi_{B_c}(0)|^2=|\Psi_{B_c^*}(0)|^2=\frac{1.642}{4\pi} {\rm GeV}^3.
\end{align}
Here, $\theta_W$ is the Weinberg angle. The $J/\psi$ wave function at the origin is extracted from its leptonic width:
\begin{equation}
\Gamma(J/\psi\to e^+e^-)=\frac{16\pi \alpha^2}{9m_c^2}|\Psi_{J/\psi}(0)|^2\left(1-4C_F\frac{\alpha_s}{\pi}\right),
\end{equation}
with $\Gamma(J/\psi\to e^+e^-)=5.55$ keV \cite{PDG}. The $B_c$ wave function is estimated by using the Buchmueller-Tye potential \cite{BTmodel}.
The two-loop strong coupling of
\begin{equation}
\frac{\alpha_s(\mu)}{4\pi}=\frac{1}{\beta_0L}-\frac{\beta_1\ln L}{\beta_0^3L^2}
\end{equation}
is employed in the NLO calculation, in which, $L=\ln(\mu^2/\Lambda_{\rm QCD}^2)$, $\beta_0=(11/3)C_A-(4/3)T_Fn_f$, $\beta_1=(34/3)C_A^2-4C_FT_Fn_f-(20/3)C_AT_Fn_f$.
We take $n_f=4$, $\Lambda_{\rm QCD}=292\ {\rm MeV}$ for $J/\psi$ and $\eta_c$ production; $n_f=5$, $\Lambda_{\rm QCD}=210\ {\rm MeV}$ for $B_c$ and $B_c^*$ production.

The NLO decay width can be expressed as
\begin{equation}
\Gamma_H^{\rm NLO}(\mu)=\frac{\alpha\alpha_s(\mu)^2}{{\rm sin}^2\theta_W}|\Psi_H(0)|^2\left(A_H+\frac{\alpha_s(\mu)}{\pi}\left(B_H+A_HC_H\ln\frac{\mu^2}{m_W^2}\right)\right),
\label{eq_nlores}
\end{equation}
where $C_H=25/6$ for $W^+\to J/\psi (\eta_c)+c+\bar{s}$, $C_H=0$ for $W^+\to \eta_c+u+\bar{d}+g$, $C_H=23/6$ for $W^+\to B_c(B_c^{*})+b+\bar{s}$.
The parameters $A_H$ and $B_H$ are independent of $\mu$, their values at different heavy quark mass are shown in Tab.\ref{tab_AB}.

\begin{table}[!htbp]
   \centering
%  \fontsize{6.5}{8}\selectfont
  \caption{The parameters $A_H$ and $B_H$ in Eq.\ref{eq_nlores}. The units of heavy quark mass and $A_H$ are suppressed for brevity, which are GeV and GeV$^{-2}$ respectively.}
  \label{tab_AB}
    \begin{tabular}{|p{2.5cm}<{\centering}|p{1.7cm}<{\centering}|p{1.7cm}<{\centering}|p{1.7cm}<{\centering}|p{1.7cm}<{\centering}|p{1.7cm}<{\centering}|p{1.7cm}<{\centering}|}
    \hline
    \multirow{2}{*}{$W^+\to Hq_i\bar{q}_j$}&
    \multicolumn{2}{c|}{$m_c=1.4$, $m_b=4.7$}&\multicolumn{2}{c|}{$m_c=1.5$, $m_b=4.9$}&\multicolumn{2}{c|}{$m_c=1.6$, $m_b=5.1$}\cr\cline{2-7}
    &$A_H$&$B_H$&$A_H$&$B_H$&$A_H$&$B_H$\cr
    \hline
    \hline
    $W^+\to J/\psi c\bar{s}$&1.04&30.5&0.846&24.2&0.695&19.6\cr\hline
   $W^+\to \eta_c c\bar{s}$&1.01&35.0&0.818&27.7&0.673&22.4\cr\hline
    $W^+\to \eta_c u\bar{d}g$&0&4.01&0&3.04&0&2.46\cr\hline
    $W^+\to B_c b\bar{s}$&0.0230&0.661&0.0201&0.569&0.0177&0.495\cr\hline
    $W^+\to B_c^* b\bar{s}$&0.0198&0.474&0.0173&0.408&0.0152&0.355\cr\hline
    \end{tabular}
\end{table}

\begin{figure}[!htbp]
\centering
\subfigure[$W^+\to J/\psi+c+\bar{s}$]{
\includegraphics[width=0.46\textwidth]{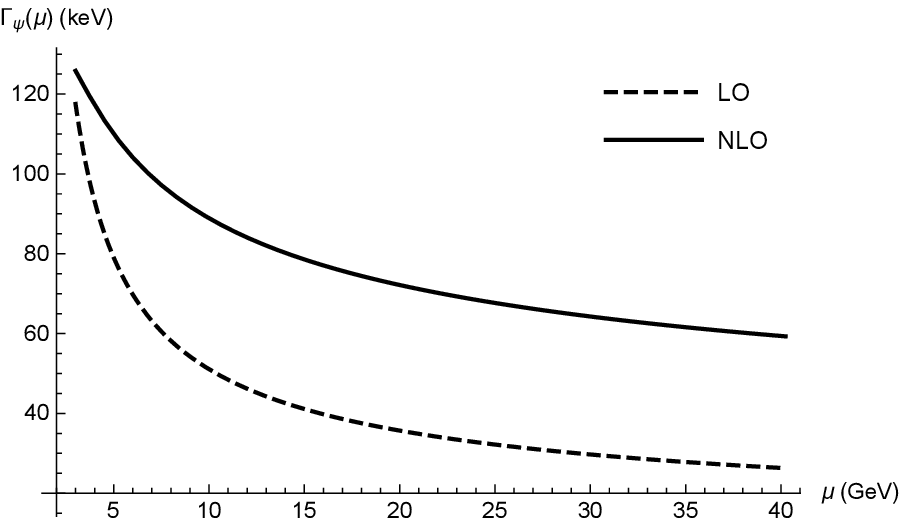}}
\subfigure[$W^+\to \eta_c+c+\bar{s}$]{
\includegraphics[width=0.46\textwidth]{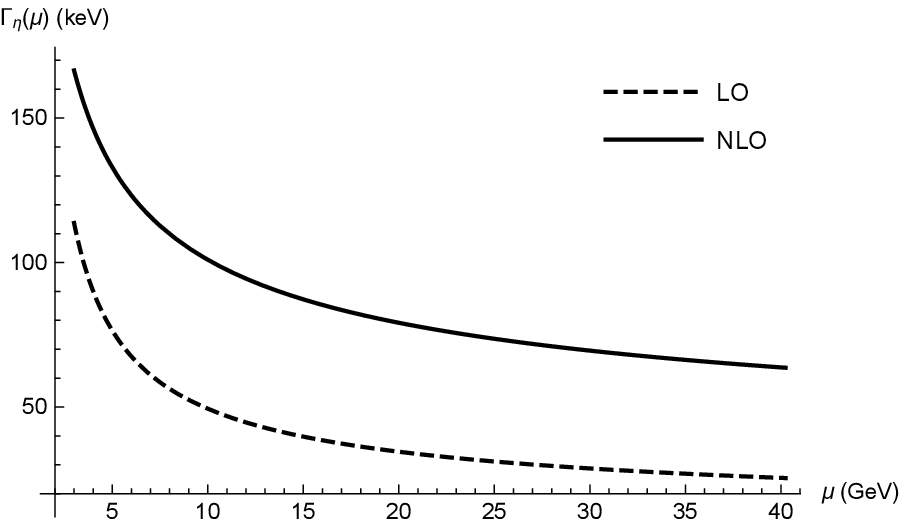}}\\
\subfigure[$W^+\to B_c+b+\bar{s}$]{
\includegraphics[width=0.46\textwidth]{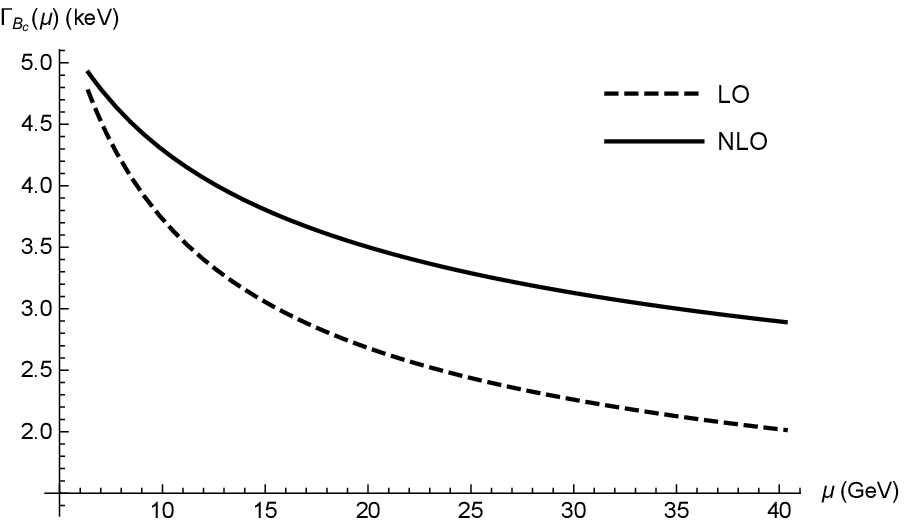}}
\subfigure[$W^+\to B_c^*+b+\bar{s}$]{
\includegraphics[width=0.46\textwidth]{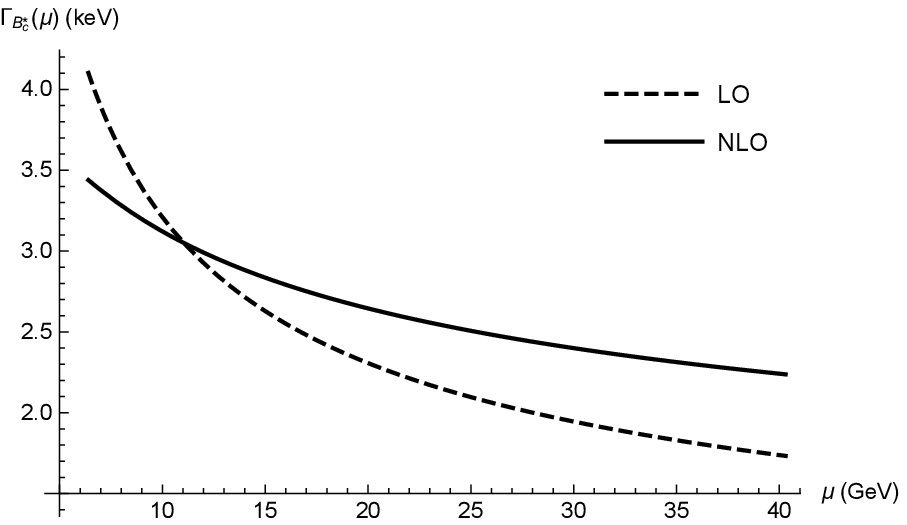}}
\caption{The LO (dashed line) and NLO (solid line) decay widths versus running renormalization scale.}
\label{fig_mudis}
\end{figure}

The decay widths are as presented in Tab.\ref{tab_tot}.
The theoretical uncertainties are estimated by varying the value of heavy quark mass and normalization scale: $m_c\in [1.4,1.6]$ GeV, $m_b\in [4.7,5.1]$ GeV, $\mu\in [m_H,\frac{m_W^2+m_H^2-m_Q^2}{2m_W}]$.
According to Ref.\cite{PDG}, total decay width for $W^+$ boson is about $2.195$ GeV, the corresponding branching fractions are then shown in Tab.\ref{tab_br}.
With NLO corrections, the theoretical uncertainties induced by heavy quark mass and normalization scale are greatly suppressed as expected.
The decay widths versus running renormalization scale at $m_c=1.5$ GeV and $m_b=4.9$ GeV are exhibited in Fig.\ref{fig_mudis}.

\begin{table}[!htbp]
\caption{Decay widths of $W^+$ inclusive decay to charmonium and $B_c(B_c^{*})$ meson. The upper bound corresponding to $m_c=1.4$ GeV, $m_b=4.7$ GeV and $\mu=m_H$, and the lower bound to $m_c=1.6$ GeV, $m_b=5.1$ GeV and $\mu=\frac{m_W^2+m_H^2-m_Q^2}{2m_W}$.}
\centering
  \begin{tabular}{p{0.8cm}<{\centering}|p{2.7cm}<{\centering}|p{2.7cm}<{\centering}|p{2.7cm}<{\centering}|p{2.7cm}<{\centering}|p{2.7cm}<{\centering}}
 \toprule[2pt]
    & $\Gamma(J/\psi c\bar{s})$(keV) & $\Gamma(\eta_c c\bar{s})$(keV) &$\Gamma(\eta_c u\bar{d}g)$(keV)&$\Gamma(B_c b\bar{s})$(keV)&$\Gamma(B_c^* b\bar{s})$(keV)     \\
    \hline
         LO & $21.6\sim 154.0$ & $20.9\sim 149.0$& - & $1.77\sim 5.62$ & $1.53\sim 4.82$ \\
 NLO & $48.3\sim 163.7$ & $51.8\sim 220.6$& $3.61\sim 45.9$ & $2.52\sim 5.82$ & $1.96\sim 4.04$ \\
 \bottomrule[2pt]
  \end{tabular}
\label{tab_tot}
\end{table}

\begin{table}[!htbp]
\caption{Branching fractions of $W^+$ inclusive decay to charmonium and $B_c(B_c^{*})$ meson.  The upper bound corresponding to $m_c=1.4$ GeV, $m_b=4.7$ GeV and $\mu=m_H$, and the lower bound to $m_c=1.6$ GeV, $m_b=5.1$ GeV and $\mu=\frac{m_W^2+m_H^2-m_Q^2}{2m_W}$.}
\centering
  \begin{tabular}{p{0.8cm}<{\centering}|p{2.8cm}<{\centering}|p{2.8cm}<{\centering}|p{2.8cm}<{\centering}|p{2.8cm}<{\centering}|p{2.8cm}<{\centering}}
 \toprule[2pt]
    & ${\rm Br}(J/\psi c\bar{s})(10^{-5})$ & ${\rm Br}(\eta_c c\bar{s})(10^{-5})$ &${\rm Br}(\eta_c u\bar{d}g)(10^{-5})$&${\rm Br}(B_c b\bar{s})(10^{-5})$&${\rm Br}(B_c^* b\bar{s})(10^{-5})$     \\
    \hline
         LO & $0.984\sim 7.02$ & $0.952\sim 6.79$& - & $0.0806\sim 0.256$ & $0.0697\sim 0.220$ \\
 NLO & $2.20\sim 7.46$ & $2.36\sim 10.05$& $0.164\sim 2.09$ & $0.115\sim 0.265$ & $0.0893\sim 0.184$ \\
 \bottomrule[2pt]
  \end{tabular}
\label{tab_br}
\end{table}

\begin{figure}[!htbp]
\centering
\subfigure[$W^+\to J/\psi+c+\bar{s}$]{
\includegraphics[width=0.46\textwidth]{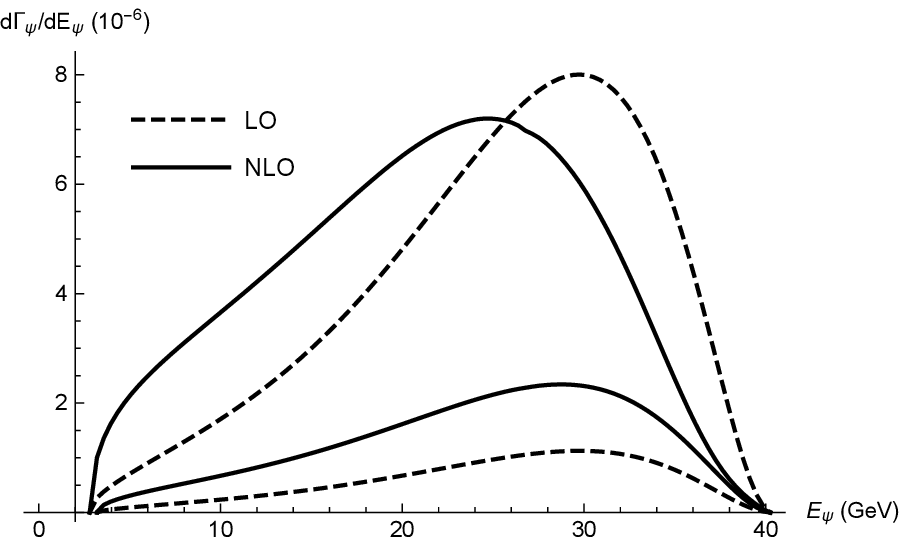}}
\subfigure[$W^+\to \eta_c+c+\bar{s}$]{
\includegraphics[width=0.46\textwidth]{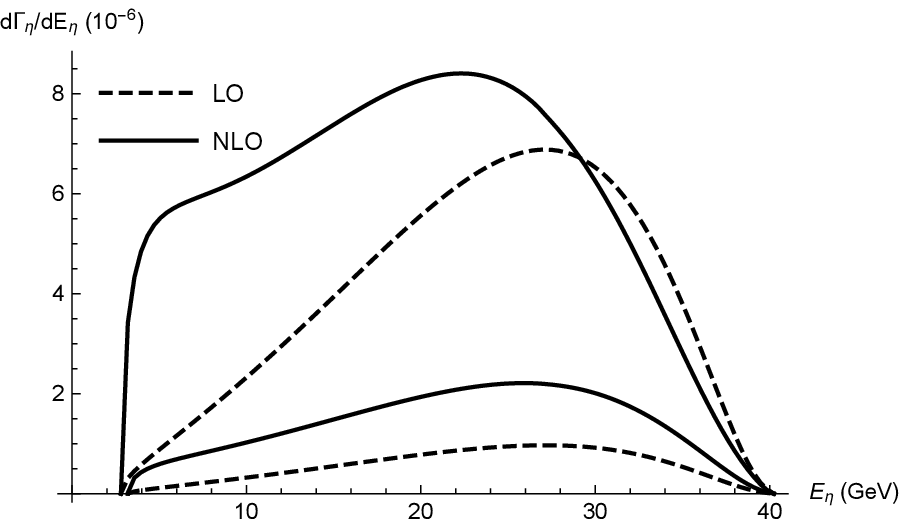}}
\subfigure[$W^+\to \eta_c+u+\bar{d}+g$]{
\includegraphics[width=0.46\textwidth]{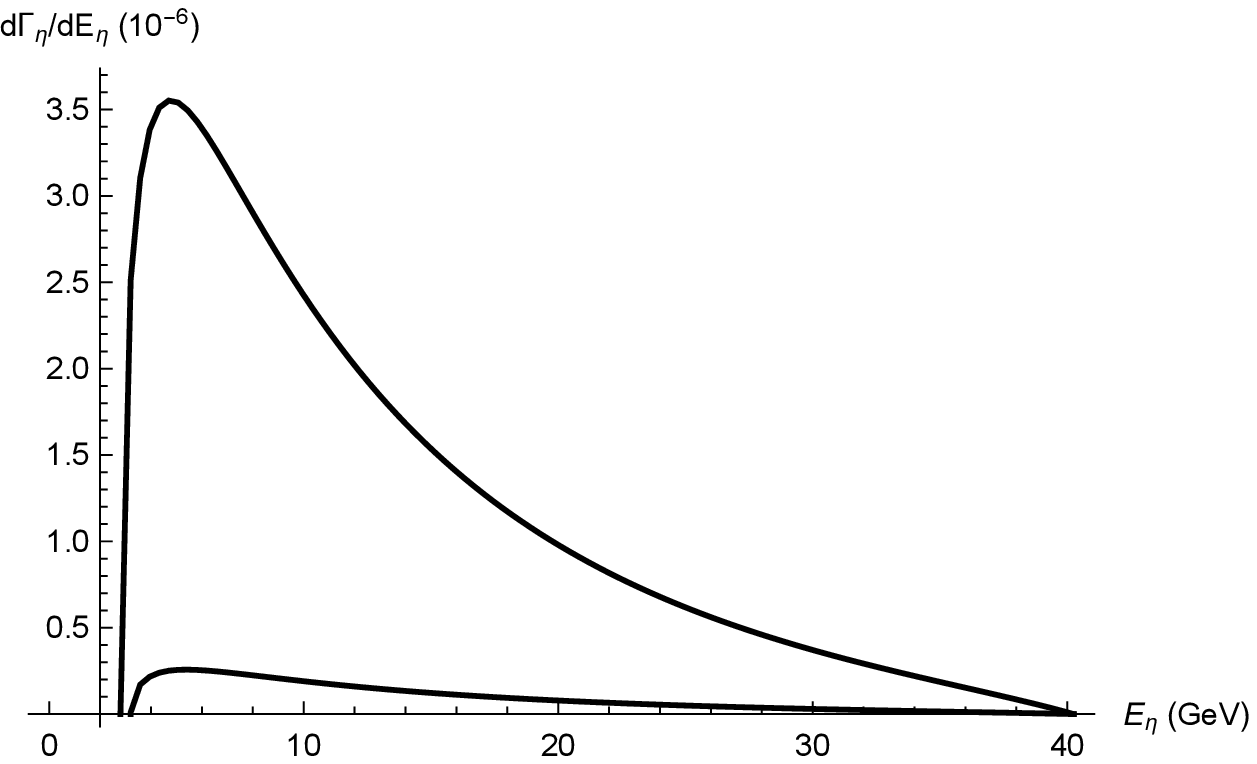}}
\subfigure[$W^+\to B_c+b+\bar{s}$]{
\includegraphics[width=0.46\textwidth]{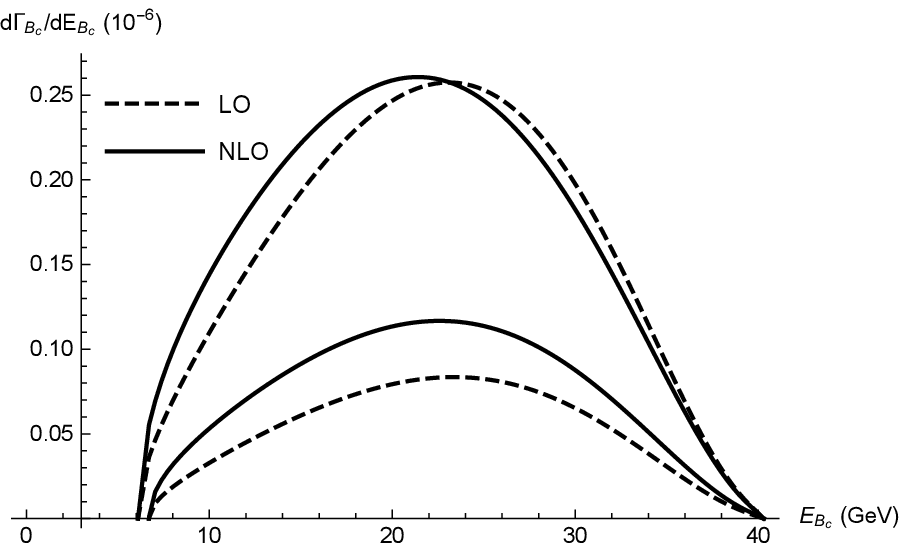}}
\subfigure[$W^+\to B_c^*+b+\bar{s}$]{
\includegraphics[width=0.46\textwidth]{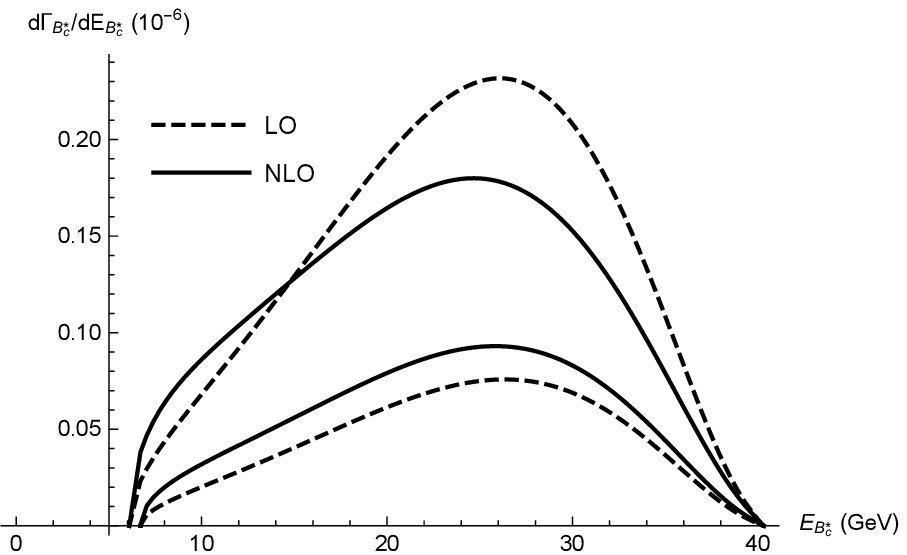}}
\subfigure[$B_c+B_c^*$]{
\includegraphics[width=0.46\textwidth]{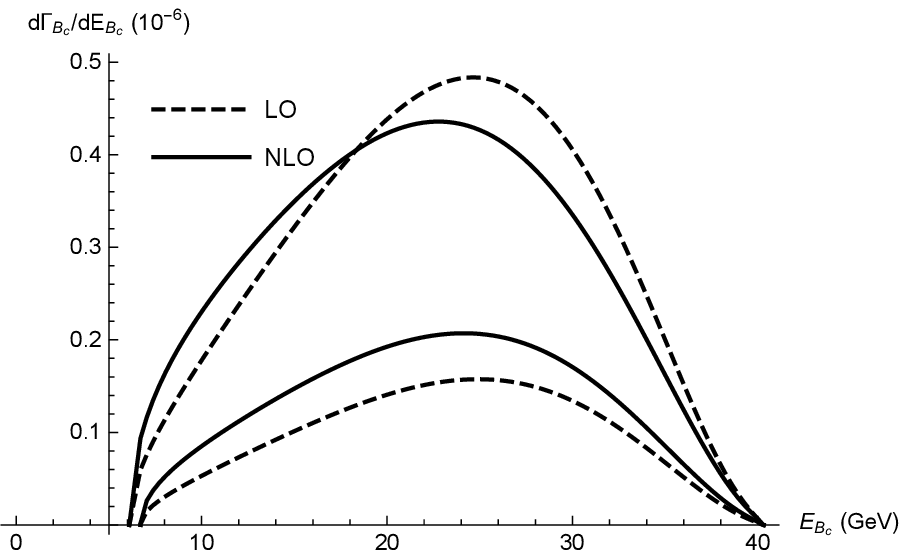}}
\caption{The charmonium and $B_c(B_c^{*})$ meson energy distribution in $W^+$ decay. The LO and NLO results are represented by double-dashed and double-solid lines, referring to the upper and lower bounds of uncertainties, respectively.}
\label{fig_Edis}
\end{figure}

The energy distribution of charmonium and $B_c(B_c^{*})$ meson are shown in Fig.\ref{fig_Edis}.
It can be seen from Fig.\ref{fig_Edis}(b) that the $\eta_c$ production rate are largely enhanced at small energy region.
This enhancement comes from the diagrams similar to Fig.\ref{fig_newpro}, except $u$ and $\bar{d}$ are replaced by $c$ and $\bar{s}$. The contribution from the gluon propagator can be estimated as:
\begin{equation}
 \displaystyle{\int_{-1}^1 d{\rm cos}\theta_{\eta g}\frac{1}{(4m_c^2+2E_\eta E_g-2|\vec{p_\eta}|E_g{\rm cos}\theta_{\eta g})^2}}\sim \frac{1}{E_gE_\eta+E_g^2+m_c^2},
\end{equation}
which explain the enhancement at small energy region.

The instantaneous luminosity of LHC reach $2.06\times 10^{34}$ cm$^{-2}$s$^{-1}$ in 2017 \cite{urllhc}.
The production cross section of $W^+$ boson at the LHC can be estimated to be $100$ nb \cite{csW},
then the number of $W^+$ events per year is about $6.5\times 10^{10}$.
Hence we can obtain about $(1.43\sim 4.85)\times 10^6$ $J/\psi$ events, $(1.64\sim 7.89)\times 10^6$ $\eta_c$ events and $(1.33\sim 2.92)\times 10^5$ $B_c$ events per year. Here, the $B_c^{*}$ feed-down to $B_c$ is taken into account.
In experiment, the $B_c$ meson can be fully reconstructed through $B_c\to J/\psi \pi^+$ decay, whose branching fraction is about $0.5\%$ \cite{BctoJpsi}.
According to \cite{PDG}, the branching ratio ${\rm B_r}(J/\psi\to l^+l^-(l=e,\mu))=12\%$, ${\rm B_r}(\eta_c\to p\bar{p})=0.15\%$,
then the numbers of $J/\psi$, $\eta_c$ and $B_c$ meson candidates per year are $(1.72\sim 5.82)\times 10^5$, $(2.46\sim 11.8)\times 10^3$ and $80\sim 175$ respectively.

\section{Summary and conclusions}
In this work we calculate the decay widths of $W^+$ to $J/\psi$, $\eta_c$ and $B_c(B_c^{*})$ mesons at the NLO QCD accuracy within the NRQCD factorization framework.
The theoretical uncertainties are estimated by varying the value of heavy quark mass and renormalization scale.
Considering there are copious $W$ data at the LHC, our results are hopefully to be tested in experiment.

Numerical calculation shows that the NLO corrections are significant, and the uncertainties in theoretical predictions with NLO corrections are greatly reduced.
Since $B_c^{*}$ alomst all decays to $B_c$, assuming $B_c$ is reconstructed through $B_c\to J/\psi \pi^+$, $J/\psi$ is reconstructed through $J/\psi\to l^+l^-(l=e,\mu)$, $\eta_c$ is reconstructed through $\eta_c\to p\bar{p}$, the numbers of $J/\psi$, $\eta_c$ and $B_c$ meson candidates per year may reach $(1.72\sim 5.82)\times 10^5$, $(2.46\sim 11.8)\times 10^3$ and $80\sim 175$ respectively at the LHC 2017 luminosity.

Note added: when this work was finished and the manuscript was finalizing, there appears a study on the web about the $B_c(B_c^{*})$ meson production in $W^+$ decay with the NLO QCD corrections \cite{WdecayZheng}. We numerically compared our results with that paper, and find that by taking the same inputs we can reproduce the Table\uppercase\expandafter{\romannumeral1} results there\footnote{In Ref.\cite{WdecayZheng}, the two-loop $\alpha_s$ is used both in the LO and NLO calculation, while in our calculation, the one-loop and two-loop $\alpha_s$ are employed respectively.}.

%%%%%%%%%%%%%%%%%%%%%%%%%%%%%%%%%%%%%%%%%%%%%%%%%%%%%%%%%%%%%%%%%%%%%%
\vspace{.7cm} {\bf Acknowledgments} \vspace{.3cm}

This work was supported in part by the Ministry of Science and Technology of the Peoples' Republic of China(2015CB856703) and by the National Natural Science Foundation of China(NSFC) under the Grants 11975236, 11635009, and 11375200.

%%%%%%%%%%%%%%%%%%%%%%%%%%%%%%%%%%%%%%%%%%%%%%%%%%%%%%%%%%%%%%%%%%%%%%%

\end{document}